\documentclass[12pt,preprint]{aastex}

\slugcomment{Article for The Astronomical Journal}

\shorttitle{Interpretation of the Helix Planetary Nebula}
\shortauthors{Djorgovski et al.}

\begin{document}


\title{Interpretation of the Helix Planetary Nebula using
Hydro-Gravitational Theory}


\author{Carl H. Gibson\altaffilmark{1}}
\affil{Departments of Mechanical and Aerospace Engineering and Scripps
Institution of Oceanography, University of California,
      San Diego, CA 92093-0411}

\email{cgibson@ucsd.edu}

\and

\author{Rudolph E. Schild}
\affil{Center for Astrophysics,
      60 Garden Street, Cambridge, MA 02138}
\email{rschild@cfa.harvard.edu}


\altaffiltext{1}{Center for Astrophysics and Space Sciences, UCSD}


\begin{abstract} Wide angle Hubble Space Telescope (HST/ACS) images of the
Helix Planetary Nebula (NGC 7293) are interpreted using the
hydro-gravitational theory (HGT) of Gibson 1996-2000 that predicts
the baryonic dark matter and interstellar medium (ISM) consist of
Mars-mass primordial-fog-particle (PFP) frozen H-He planets.  The new ACS
images confirm and extend the O'Dell and Handron 1996 WFPC2 images showing
thousands of cometary globules, which we suggest are cocoons of PFP and
Jupiter frozen-gas-planets evaporated by powerful beamed radiation from the
hot central white dwarf and its companion.  The atmosphere mass of the
largest cometary globules is
$\approx 3 \times 10^{25}$ kg with spacing
$\approx 10^{14}$ m, supporting the prediction of HGT  that the mass
density of the ISM in Galaxy star forming regions should match
the large baryonic primordial value at the time of first structure formation
($10^{12}$ s), with
$\rho \approx (3-1) \times 10^{-17}$ kg m$^{-3}$.   
\end{abstract}




\section{Introduction}

 By coincidence, the direction opposite to the peak Leonid meteoroid flux
in November 2002 includes the closest planetary nebula (PNe) to the earth
(150-200 pc $ \approx 5 \times 10^{18}$ m) Helix (NGC 7293), so that the
Hubble Helix team of volunteers could devote a substantial fraction of the
14 hour Leonid stand-down period taking photographs with the full array of
HST cameras including the newly installed wide angle Advanced Camera for
Surveys (ACS).  As the closest PNe to earth with one of the hottest known
central white dwarf stars (120,000 K) and close (dMe) X-ray 
companion
\citep{gue01} to powerfully beam radiation and plasma into their
surroundings, Helix provides an ideal laboratory to test predictions
of hydro-gravitational theory \citep{gib96} and a quasar microlensing
interpretation
\citep{sch96} that the ISM of star forming regions of the
Galaxy are dominated by primordial densities of baryonic dark matter in the
form of frozen H-He primordial-fog-particle (PFP)  planets from which all
stars and planets have formed.  According to HGT, most Galactic star
formation occurs in primordial proto-globular-star-cluster (PGC) clumps of
PFPs with baryonic mass density
$\approx (3-1)
\times 10^{-17}$ kg m$^{-3}$ reflecting the large mass density existing at
the time of the first gravitational structure formation 30,000 years after
the big bang.  

Most of the $\approx$million
PGCs in the Galaxy have diffused away from the central core to form the dark
matter inner halo.  From the HST/ACS images of the Tadpole merger, the
inner halo is homogeneous and very large. Its 130 kpc radius is
revealed by young globular clusters and stars triggered into formation by
tidal forces and radiation
 from the merging galaxy \citep{gs03a}.  Galactic star
formation occurs in the disk.   From HGT the disk is formed by accreting
PGCs, so the interstellar medium of the disk should be that of a PGC;
that is, trillions of frozen  PFPs in metastable equilibrium with their
evaporated gas at the high average baryonic mass density of the time of
first structure, $\rho_B
\approx 10^{4}
\times
\langle
\rho_{G} \rangle$, where the average density of the Galaxy $ \langle
\rho_{G} \rangle \approx 10^{-21}$ kg m$^{-3}$.  From HGT it is no accident
that the outer planets of the solar system are large and gassy and that the
hundreds of extra-solar planets discovered are also large Jovians.   Hot,
central, PNe white dwarfs (generically with companion stars) form plasma
beams and radiation that evaporate and photo-ionize surface layers of their
invisible surrounding ISM of PFP frozen-gas-planets and their accretional
Jovian descendants (JPPs) to large sizes and brightnesses so that their
evaporated gas cocoons can be imaged and properties measured for comparison
with theory.  In this paper we compare the new HST/ACS Helix observations,
plus other PNe observations, with hydro-gravitational theory  and
previously proposed explanations of cometary globule and planetary nebula
formation.

According to HGT, gravitational structure formation began in the plasma
epoch soon after mass dominated energy at time 25,000 years with
gravitational fragmentation of the plasma into $\approx 10^{47}$ kg
proto-supercluster mass (baryonic and non-baryonic) clouds that further
fragmented  to proto-cluster and then
$\approx 10^{42}$ kg proto-galaxy mass (baryonic) clouds just before the
plasma to gas transition at 300,000 years.  At that time, fragmentation
scales decreased dramatically when 
the kinematic viscosity decreased from  $\nu
\approx 10^{25}$ m$^2$ s$^{-1}$ for the primordial plasma to
$\nu \approx 10^{13}$ m$^2$ s$^{-1}$ for the  H-He gas upon photon
decoupling
\citep{gib00}.  An important prediction of HGT is
that gravitational fragmentation occurred in the primordial plasma at length
scales much smaller than the Jeans 1902 acoustic gravitational scale (Table
1) and that the more massive non-baryonic dark matter (possibly neutrinos)
diffused into the voids to reduce the gravitational driving force. 
Thus, the non-baryonic dark matter did not clump to form cold
dark matter (CDM) halos that further clumped to form potential wells for
galaxy formation as predicted by the standard (but erroneous and
misleading) cold-dark-matter-hierarchical-clustering cosmological (CDMHCC)
model.   Gravitational instability and structure formation begins from
non-acoustic density nuclei at the smaller  scales permitted by the smaller
diffusivity of baryonic matter.  Such density nuclei are absolutely
unstable to gravity, and immediately grow or decrease in magnitude
depending on fluid forces at the largest Schwarz scale, Table 1.

Proto-galaxies formed just before photon decoupling at $10^{13}$ s retain
the baryonic density existing at the earlier time $10^{12}$ s of first
structure ($\rho_B = (3-1) \times 10^{-17}$ kg m$^{-3}$) as a fossil of that
event, and fragment simultaneously at the Jeans scale and the viscous
Schwarz scale determined by the fossilized rate-of-strain ($\gamma
\approx 10^{-12}$ s$^{-1}$) of the plasma at the time of first structure to
form proto-globular-star-clusters (PGCs) and primordial-fog-particles
(PFPs).  Table 1 lists the critical length scales of primordial
self-gravitational structure formation and their physical significance. 
Table 2 lists some of the acronyms used to describe the process of
structure formation by HGT.  A full discussion of HGT with comparisons to
recent observations is available elsewhere
\citep{gs03b}.

In the following we briefly review hydro-gravitational theory and some
of the supporting evidence, and compare the predictions of HGT with respect
to planetary nebula and star formation with standard
models and the observations.  Finally, some conclusions are offered.

\section{Theory}

\subsection {Hydro-gravitational structure formation}

Standard CDMHC cosmologies are based on ill-posed, over-simplified fluid
mechanical equations, an inappropriate assumption that the fluid is
collisionless, and the ``Jeans swindle'' to achieve a solution.  The Jeans
1902 theory neglects non-acoustic density fluctuations, viscous forces,
turbulence forces,  particle collisions, and the effects of diffusion on
gravitational structure formation, all of which can be crucially important
in some circumstances where astrophysical structures form by self gravity. 
Jeans did linear perturbation stability analysis (neglecting turbulence) of
Euler's equations (neglecting viscous forces) for a completely  uniform
ideal gas with density
$\rho$ only a function of pressure (the barotropic assumption) to reduce
the problem of self-gravitational instability to one of gravitational
acoustics.  To satisfy Poisson's equation for the gravitational potential
of a collisionless ideal gas, Jeans assumed the density $\rho$ was zero in a
maneuver appropriately known as the ``Jeans swindle''.    The only critical
wave length for gravitational instability with all these questionable
assumptions is the Jeans acoustical length scale
$L_J$ where
\begin{equation}  L_J \equiv V_S/(\rho G)^{1/2}
\gg (p/\rho^2 G)^{1/2} \equiv L_{JHS} ,
\label{eq1}
\end{equation}
  $G$ is Newton's gravitational constant and
$V_S \approx (p/\rho)^{1/2}$ is the sound speed.  

The related Jeans hydrostatic length scale $L_{JHS} \equiv
[p/\rho^2G]^{1/2}$ is given in the equation and has been 
misinterpreted by Jeans 1902 and others as an indication that
pressure itself somehow prevents gravitational condensation on scales
smaller than
$L_J$.  Although the two scales appear to be equal they are not.  The
ratio $h = p/\rho$ in $L_{JHS}$ is the stagnation enthalpy for a
condensation streamline and is initially zero from Bernoulli's equation. 
Non-acoustic density extrema are absolutely unstable to gravitational
structure formation.  Minima trigger  voids and maxima trigger condensates
at all scales not stabilized by turbulent forces, viscous forces, other
forces, or diffusion.   The Jeans acoustic scale
$L_J$ is the size for which pressure can equilibrate acoustically without
temperature change in an ideal gas undergoing self gravitational collapse
or void formation, smoothing away all pressure forces and all pressure
resistance to self gravity.  The Jeans hydrostatic scale
$L_{JHS}$ is the size of a fluid blob for which irreversibilities such as
frictional forces or thermonuclear heating have achieved a hydrostatic
equilibrium between pressure and gravitation. 
$L_{JHS}$ is generically much smaller than
$L_J$ and has no physical significance until gravitational condensation has
actually occurred and a hydrostatic equilibrium has been achieved.  

 When gas
condenses on a non-acoustic density maximum due to self gravity a variety
of results are possible.  If the amount is large a turbulent maelstrom,
superstar and black hole will appear.  If the amount is small a gas
planet, brown dwarf or star will form in hydrostatic equilibrium as the
gravitational potential energy is converted to heat by turbulent friction
and radiated.  The pressure force
$F_P
\approx p \times L^2$ matches the gravitational force of the star or gas
planet at $F_G \approx
\rho^2 G L^4$ at the hydrostatic Jeans scale
$L_{JHS}$.  Pressure $p$ is determined by a complex mass-momentum-energy
balance of the fluid flow and ambient conditions.  A gas with uniform
density is absolutely unstable to self gravitational structure formation on
non-acoustic density perturbations at scales larger and smaller than $L_J$
and is unstable to acoustical density fluctuations on scales larger than
$L_J$
\cite{gib96}. Contrary to the common misconception, pressure and temperature
cannot prevent structure formation on scales larger or smaller than
$L_J$.

Density fluctuations in fluids are not barotropic  as assumed by Jeans 1902
except rarely in small regions for short times near powerful sound
sources.  Density fluctuations that triggered the first gravitational
structures in the primordial fluids of interest were likely non-acoustic
(non-barotropic) density variations from turbulent mixing of temperature or
chemical species  concentrations produced by the big bang
\cite{gib01} as shown by turbulent signatures \cite{bs02} in the cosmic
microwave background temperature anisotropies.  From Jeans' theory without
Jeans' swindle, a gravitational condensation on an acoustical density
maximum rapidly becomes a non-acoustical density maximum  because the
gravitationally accreted mass retains the (zero) momentum of the motionless
ambient gas.  The Jeans 1902 analysis was ill posed because it failed to
include non-acoustic density variations as an initial condition.

Fluids with non-acoustic density fluctuations are continuously in a state
of structure formation due to self gravity unless prevented by diffusion or
fluid forces
\cite{gib96}.  Turbulence or viscous forces can dominate gravitational
forces at small distances from a point of maximum or minimum density to
prevent gravitational structure formation, but gravitational forces will
dominate turbulent or viscous forces at larger distances to cause
structures if the gas or plasma does not diffuse away faster than it can
condense or rarify due to gravity.  The erroneous concepts of pressure
support and thermal support are artifacts of the erroneous Jeans criterion
for gravitational instability.  Pressure forces could not prevent
gravitational structure formation in the plasma epoch because pressures
equilibrate in time periods smaller that the gravitational free fall time
$(\rho G)^{-1/2}$ on length scales smaller than the Jeans scale $L_J$, and
$L_J$ in the primordial plasma was larger than the Hubble scale of
causal connection $L_J > L_H = ct$, where $c$ is light speed and $t$ is
time.  Therefore, if gravitational forces exceed viscous and turbulence
forces in the plasma epoch at Schwarz scales $L_{ST}$ and $L_{SV}$ smaller
than
$L_H$ (Table 1) then gravitational structures will develop, independent of
the Jeans criterion.  Only a very large diffusivity ($D_B$) could interfere
with structure formation in the plasma.  Diffusion prevents gravitional
clumping of the non-baryonic dark matter (cold or hot) in the plasma epoch
because
$D_{NB} \gg D_B$ and $(L_{SD})_{NB} \gg L_H$.

Consider the gravitational response of a large body of
uniform density gas to a sudden change at time  $t=0$ on scale
$L \ll L_J$ of a rigid mass $M(t)$ at the center, either a cannonball or
vacuum beach ball depending on whether $M(0)$ is positive or negative.
Gravitational forces will cause all the surrounding gas to accelerate slowly
toward or away from the central mass perturbation.  The radial velocity
$v_r = -GM(t)t r^{-2}$ by integrating the radial gravitational acceleration
and the central mass increases at a rate $dM(t)/dt = -v_r 4 \pi r^2 \rho =
4 \pi \rho G M(t)t$.  Separating variables and integrating gives $M(t) =
M(0) exp(2 \pi \rho G t^2)$.  Nothing much happens for time periods less
than the gravitational free fall time $t_G = (\rho G)^{-1/2}$ except for a
gradual build up or depletion of the gas near the center where the
hydrostatic pressure changes are concentrated at the Jeans hydrostatic
scale $L_{JHS} \ll L_J$.  At
$t = 0.43 t_G$ the mass ratio $M(t)/M(0)$ for $r < L$ has increased by only
a factor of 2.7, but goes from 534 at $t = t_G$ to  $10^{11} $ at
$t = 2 t_G$ during the time it would take for an acoustic signal to reach a
distance $L_J$.  Pressure support and the Jeans 1902 criterion clearly
fail in this exercise.

 The diffusion velocity is
$D/L$ for diffusivity
$D$ at distance $L$ and the gravitational velocity is $L (\rho G)^{1/2}$.
The two velocities are equal at the diffusive Schwarz length scale
\begin{equation} L_{SD} \equiv [D^2 / \rho G]^{1/4}.\end{equation}  Weakly
collisional particles such as the hypothetical cold-dark-matter (CDM)
material cannot possibly form clumps, seeds, halos, or  potential wells for
baryonic matter collection because the CDM particles have large diffusivity
and will disperse, consistent with observations \cite{sa02}.  Diffusivity
$D 
\approx V_p \times L_c$, where $V_p$ is the particle speed and $L_c$ is the
collision distance. Because weakly collisional particles have large
collision distances with large diffusive Schwarz lengths the non-baryonic
dark matter (possibly neutrinos) is the last material to fragment by self
gravity and not the first as assumed by CDM cosmologies.  The first
structures occur as proto-supercluster-voids in the baryonic plasma
controlled by viscous and weak turbulence forces, independent of
diffusivity  ($D \approx \nu$).  The CDM seeds postulated as the basis of
CDMHCC never happened because $(L_{SD})_{NB}
\gg ct$ in the plasma epoch.  Because CDM seeds and halos never happened,
hierarchical clustering of CDM halos to form galaxies and their clusters
never happened.

Cold dark matter was invented to explain the observation that gravitational
structure formed early in the universe that should not be there from the
Jeans 1902 criterion that forbids structure in the baryonic plasma because
$(L_J)_B > L_H$ during the plasma epoch (where sound speed approached light
speed
$V_S =c/\surd 3$). In this CDM cosmology, non-baryonic particles with rest
mass sufficient to be non-relativistic at their time of decoupling are
considered ``cold'' dark matter, and are assumed to form permanent, cohesive
clumps in virial equilibrium that can only interact with matter and other
CDM clumps gravitationally.  This assumption that CDM clumps are cohesive
is questioned here as unnecessary and unrealistic.  Numerical simulations
of large numbers of such cohesive CDM clumps show a tendency for the
clumps to clump further due to gravity to form ``halos'', justifying the
cold dark matter hierarchical clustering cosmology (CDMHCC) where the
baryonic matter eventually falls into the gravitational potential wells of
the CDM halos, cools off, and forms the observed stars and structures.  As
we have seen, CDMHCC is not necessary since the Jeans criterion is
incorrect and baryons can begin gravitational structure formation during the
plasma epoch.  Furthermore, CDMCC is not possible because the nearly
collisionless non-baryonic matter would not clump or form massive halos as
assumed but would diffuse away. 

 Clumps of
collisionless or collisional CDM would either form black holes or thermalize
in time periods of order the gravitational free fall time $(\rho G)^{-1/2}$
because the particles would gravitate to the center of the clump by core
collapse where the density would exponentiate, causing double and triple
gravitational interactions or particle collisions that would thermalize the
velocity distribution and trigger diffusional evaporation.  For collisional
CDM, consider a spherical clump of perfectly cold CDM with mass
$M$, density
$\rho$, particle mass $m$ and collision cross section $\sigma$.  The clump
collapses in time $(\rho G)^{-1/2}$ to density $\rho_c =
(m/\sigma)^{3/2}M^{-1/2}$ where collisions begin and the velocity
distribution thermalizes.  Particles with velocities greater than the
escape velocity $v \approx 2MG/r$ then diffuse away from the clump,
where
$r=(M/\rho)^{1/3}$ is the initial clump size.  For typically
considered CDM clumps of mass $\approx 10^{36}$ kg and CDM particles more
massive than 
$10^{-24}$ kg (WIMPs with $\sigma \approx 10^{-42}$ m$^2$ small
enough to escape detection) the density from the expression would require a
collision scale smaller than the clump Schwarzschild radius so that such
CDM clumps would collapse to form black holes.  Less massive motionless CDM
particles collapse to diffusive densities smaller than the
black hole density, have collisions, thermalize, and diffuse away.  From the
outer halo radius size measured for galaxy cluster halos it is possible to
estimate the non-baryonic dark matter particle mass to be of order
$10^{-35}$ kg (10 ev) and the diffusivity to be $\approx 10^{-30}$ m$^2$
s$^{-1}$
\citep{gib00}.  Thus, CDM clumps are neither necessary nor physically
possible, and are ruled out by observations
\citep{sa02}.  It is  recommended that the CDMHCC scenario for
structure formation and cosmology be abandoned.

The baryonic matter is subject to large viscous forces, especially in the
hot primordial plasma and gas states  existing when most gravitational
structures first formed \citep{gib00}.  The viscous forces per unit  volume
$\rho
\nu
\gamma L^2$ dominate gravitational forces $\rho^2 G L^4$ at small scales,
where
$\nu$ is the kinematic viscosity and $\gamma$ is the rate of strain of the
fluid.  The forces match at the viscous Schwarz length
\begin{equation} L_{SV} \equiv (\nu \gamma /
\rho G)^{1/2},\end{equation}
  which is the smallest size for self gravitational condensation or void
formation in such a flow.  Turbulent forces may permit larger mass
gravitational structures to develop; for example, in thermonuclear
maelstroms at galaxy cores to form central black holes.  Turbulent forces
$\rho
\varepsilon^{2/3} L^{8/3}$ match gravitational forces at the turbulent
Schwarz scale
\begin{equation}L_{ST} \equiv \varepsilon ^{1/2}/(\rho
G)^{3/4},\end{equation}
  where $\varepsilon$ is the viscous dissipation rate of the turbulence. 
Because in the primordial plasma the viscosity and diffusivity are identical
and the rate-of-strain $\gamma$ is larger than the free-fall frequency
$(\rho G)^{1/2}$, the viscous and turbulent Schwarz scales
$L_{SV}$ and 
$L_{ST}$ will be larger than the diffusive Schwarz scale $L_{SD}$, from
(2), (3) and (4).  

The criterion for structure formation in the plasma epoch is that
both $L_{SV}$ and $L_{ST}$ become less than the horizon scale $L_H = ct$. 
Reynolds numbers in the plasma epoch were near critical, with  $L_{SV}
\approx L_{ST}$.  From $L_{SV}< ct$, gravitational structures first formed
when
$\nu$ first decreased to values less than radiation dominated values $c^2 t
$ at time
$t
\approx 10^{12}$ seconds \cite{gib96}, well before $10^{13}$ seconds which
is the time of plasma to gas transition (300,000 years).  Because the
expansion of the universe inhibited condensation but enhanced void
formation in the weakly turbulent plasma, the first structures were
proto-supercluster-voids in the baryonic plasma. At
$10^{12}$ s 
\begin{equation} (L_{SD})_{NB} \gg L_{SV} \approx L_{ST} \approx 5 \times
L_K \approx L_H = 3
\times 10^{20} \rm m \gg L_{SD}, \end{equation}  where $(L_{SD})_{NB}$
refers to the non-baryonic component and $L_{SV}$,
$L_{ST}$, $L_{K}$, and $L_{SD}$ scales refer to the baryonic (plasma)
component.

As proto-supercluster mass plasma fragments formed, the voids filled with
non-baryonic matter by diffusion, thus inhibiting further structure
formation by decreasing the gravitational driving force.  The baryonic mass
density
$\rho
\approx 2
\times  10^{-17}$ kg/$\rm m^3$ and rate of strain
$  \gamma \approx 10^{-12}$ $\rm s^{-1}$ were preserved as hydrodynamic
fossils within the proto-supercluster fragments and within proto-cluster
and proto-galaxy objects resulting from subsequent fragmentation as the
photon viscosity and
$L_{SV}$ decreased prior to the plasma-gas transition and photon decoupling
\cite{gib00}.  As shown in Eq. 5, the Kolmogorov scale $L_K \equiv [\nu^3
/\varepsilon ]^{1/4}$ and the viscous and turbulent Schwarz scales at the
time of first structure matched the horizon scale $L_H
\equiv ct \approx 3 \times 10^{20}$ m, freezing in the density,
strain-rate, and spin magnitudes and directions of the subsequent
proto-cluster and proto-galaxy fragments of proto-superclusters.  Remnants
of the strain-rate and spin magnitudes and directions of the weak
turbulence at the time of first structure formation are forms of fossil
vorticity turbulence
\cite{gib99}.

The quiet condition of the primordial gas is  revealed by measurements of
temperature fluctuations of the cosmic microwave background  radiation that
show an average $\delta T/T \approx 10^{-5}$ much too small for any 
turbulence to have existed at that time of plasma-gas transition ($10^{13}$
s).  Turbulent plasma motions are strongly damped by buoyancy forces at
horizon scales after the first gravitational fragmentation time 
$10^{12}$ s.  Viscous forces in the plasma are inadequate to explain the
lack of primordial turbulence ($\nu$
$ \ge 10^{30}$ m$^2$ s$^{-1}$ is required but, after $10^{12}$ s, $\nu \le 4
\times 10^{26}$, Gibson 2000). The observed lack of plasma turbulence
proves that large scale buoyancy forces, and therefore self gravitational
structure formation, must have begun in the plasma epoch.

The gas temperature, density, viscosity, and rate of strain are all
precisely known at transition, so the gas viscous Schwarz  mass
$L_{SV}^3 \rho$ is easily calculated to be about
$10^{24}$ kg, the mass of a small planet, or about
$10^{-6} M_{\sun}$, with uncertainty about a factor of ten.  From HGT, soon
after the cooling primordial plasma turned to gas at $10^{13}$ s (300,000
yr), the entire baryonic universe condensed to a fog of planetary-mass
primordial-fog-particles (PFPs) that preventing collapse at the Jeans mass. 
These gas-cloud objects gradually cooled, formed H-He rain, and eventually
froze solid to become the baryonic dark matter and the basic material of 
construction for stars and everything else, about $30 \times 10^{6}$ rogue
planets per star.

The Jeans mass $L_J^3 \rho$ of the primordial gas at transition was about
$10^6 M_{\sun}$ with about a factor of ten uncertainty, the mass of a
globular-star-cluster (GC).  Proto-galaxies fragmented at the PFP scale but
also at this proto-globular-star-cluster PGC scale
$L_J$, although not for the reason given by the Jeans 1902 theory.  Density
fluctuations in the gaseous proto-galaxies were absolutely unstable to 
void formation at all scales larger than the viscous Schwarz scale
$L_{SV}$.  Pressure can  only remain in equilibrium with density without
temperature changes in a gravitationally expanding void on scales smaller
than the Jeans scale.  From the second law of thermodynamics, rarefaction
wave speeds that develop as density  minima expand due to gravity to form
voids are limited to speeds less than the sonic velocity.  Cooling could
therefore occur and be compensated by radiation in the otherwise isothermal
primordial gas when the expanding voids approached the Jeans scale.
Gravitational fragmentations of  proto-galaxies were then accelerated by
radiative heat transfer to these cooler regions, resulting in fragmentation
at the Jeans scale and isolation of proto-globular-star-clusters (PGCs)
with the primordial-gas-Jeans-mass.  

These
$10^{36}$ kg PGC objects were not able to collapse from their own self
gravity  because of their internal fragmentation at the viscous Schwarz
scale to form $10^{24}$ kg PFPs. The fact that globular star clusters have
precisely the same density and primordial-gas-Jeans-mass from galaxy to
galaxy proves they were all formed simultaneously soon after the time of
the plasma to gas transition $10^{13}$ s.  The gas has never been so
uniform since, and no mechanism exists to recover such a high density, let
alone such a high uniform density, as the fossil turbulent density value
$\rho \approx 2 \times 10^{-17}$ kg/$\rm m^3$.  Young globular cluster
formation in BDM halos in the Tadpole, Mice, and Antennae galaxy mergers
\citep{gs03a} show that dark PGC clusters of PFPs are remarkably stable
structures, persisting without disruption or star formation for more than
ten billion years.

\subsection{Observational evidence for PGCs and PFPs}

Searches for point mass objects as the dark matter by looking
for microlensing of stars in the bulge and the Magellanic clouds
have detected only about 20$\%$ of the expected
amount, leading to claims by the MACHO/OGLE/EROS consortia that this
form of dark matter has been observationally excluded \citep{alc98}. 
These studies have assumed a uniform rather than clumped
distribution for the ``massive compact halo objects'' (MACHOs) and used
sampling frequencies appropriate for stellar rather than small planetary
mass objects.  Furthermore, since the PFPs within PGC clumps must
accretionally cascade over a million-fold mass range to produce JPPs and
stars their statistical distribution becomes an intermittent lognormal that
profoundly affects the sampling strategy and microlensing data
interpretation and makes the exclusion of PFP mass objects as the baryonic
dark matter (BDM) of the Galaxy premature
\citep{gs99}.  Recent OGLE campaigns focusing on planetary mass to brown
dwarf mass objects have revealed 121 transiting and orbiting candidates,
some with orbits less than one day indicating the end of the PFP
accretional cascade predicted by HGT as the mechanism of star formation
\citep{uda03}.

Evidence that planetary mass objects dominate the BDM in galaxies has been
gradually accumulating and has been reviewed \citep{gs03b}. 
Cometary knot candidates for PFPs and JPPs appear whenever hot events 
like white dwarfs, novas, plasma jets, Herbig-Haro objects, and supernovas
happen, consistent with the prediction of HGT that the knots reveal Jovian
planets that comprise the BDM, as we see for the planetary nebulae in the
present paper.  However, the most convincing evidence for our hypothesis,
because it averages the dark matter over much larger volumes of space, is
provided by one of the most technically challenging areas in astronomy; that
is, quasar microlensing
\citep{sch96}.  Several years and many dedicated observers were required to
confirm the Schild measured time delay of the Q0957 lensed quasar images so
that the twinkling of the subtracted light curves could be confirmed and the
frequency of twinkling interpreted as evidence that the dominant point mass
objects of the lensing galaxy were of small planetary mass.  By using
multiple observatories around the Earth it has now been possible to
accurately establish the Q0957 time delay at
$417.09  \pm 0.07$ days (Colley et al. 2002, 2003).  With this
unprecedented accuracy a statistically significant microlensing event of
only 12 hours has now been detected
\citep{ColS03} indicating a PFP with Moon-mass only $7.4
\times 10^{22}$ kg.  An additional microlensing system has been
observed (Schechter et al. 2003) and confirmed, and its time delay measured
(Ofek and Maoz 2003).  To attribute the microlensing to stars rather than
planets required Schechter et al. to propose relativistic knots in the
quasar.  An additional four lensed quasar systems with  measured time
delays show monthly period microlensing that support dominant BDM objects
in the lens with planetary mass (Burud et al. 2000, 2002; Hjorth et al.
2002).

Flux anomalies in four-image gravitational lenses have been interpreted as
evidence for the dark matter substructure predicted by CDM halo models
\citep{dal02}, but the anomalies can also be taken as evidence for
concentrations of baryonic dark matter such as PGCs, especially when the
images are found to twinkle with frequencies consistent with the existence
of planetary mass objects.  Further evidence that the planetary objects
causing the high frequency twinkling are clumped as PGCs is provided by
evidence that the HE1104 system \citep{sche03} has a damped Lyman alpha
lensing system (DLA $\equiv$ neutral hydrogen column density larger than
$10^{24.3}$ m$^{-2}$), which is therefore a PGC candidate from the
evidence of gas and planets.  Active searches are underway for lensed DLAs
with planetary frequency twinkling that can add to this evidence of PGCs. 
Perhaps the most remarkable evidence suggesting galaxy inner halos consist
mostly of baryonic PGC-PFP clumps are the recent HST/ACS images of aligned
rows of YGCs precisely tracking the merging galaxy in the
Tadpole system
\citep{gs03a}.

Numerous YGCs are also seen in the fragmenting galaxy cluster Stephan's
Quintet \citep{gs03c}.  The mysterious red shifts of the dense-cluster
support the HGT model of sticky beginnings of the cluster in the plasma
epoch, where viscous forces of the baryonic dark matter halo of the cluster
have inhibited the final breakup due to the expansion of the universe to
about 200 million years ago and reduced the transverse velocities of the
galaxies to small values so that they appear aligned in a thin pencil by
perspective.  Close alignments of QSOs with bright galaxies (suggesting
intrinsic red shifts) have been noted for many years \citep{hoy00} that can
more easily be explained by HGT without requiring new physics.

\subsection{Planetary Nebula formation}
\subsubsection{The standard model}

According to the standard model of planetary nebula formation, an ordinary
star like the sun eventually burns most of its hydrogen and helium to form a
hot, dense, carbon core.  In its last stages the remaining H-He fuel forms
a progressively less dense atmosphere that expands from $ \approx 10^{9}$ m
to
$10^{11}$ m or more to form a cool 3000 K red giant cocoon around the
carbon center. This neutral atmosphere of the red giant with density
$\approx 10^{-17}$ kg m$^{-3}$ is eventually expelled by dynamical and photon
pressures when the hot
$\approx 10^{5}$ K, dense $\approx 10^{10}$ kg m$^{-3}$, carbon core is
exposed as a white dwarf star with no source of fuel unless accompanied by
a donor companion.  The density of this
$10^{16}$ kg atmosphere at a distance of $3 \times 10^{15}$ m corresponding
to the inner Helix radius is only $3.7 \times 10^{-31}$ kg m$^{-3}$, which
is $2.7
\times 10^{16}$ times smaller than the observed density of the cometary
knots.  Larger stars up to $5 M_{\sun}$ form white dwarfs and have complex
histories with super-wind Asymptotic Giant Branch (AGB) periods where many
solar masses are expelled into the ISM \citep{bus99}.  At most this could
bring the density within a factor of $10^3$ of that observed in the knots. 
It has been proposed in versions of the standard model that either the
cometary knots are produced by shock wave instabilities and are somehow
ejected by the central stars
\citep{vis94} or the 
photo-ionized inner surface of the denser ejected atmosphere forms
Rayleigh-Taylor instabilities so that fingers like
dripping paint break into the cometary globules and radial wakes
observed
\citep{cap73}.

Several problems exist for this standard model.  The huge masses of the
observed planetary nebulae H-He gasses are much larger and richer in other
species and dust than one would expect to be expelled as stellar winds or
cometary bullets during any reasonable model of star evolution, where most
of the star's H-He fuel should be converted by thermonuclear fusion to
carbon in the core before the star dies.  Instead, more than a solar mass
of gas and dust is found in the nebular ring of Helix, with a dusty
H-He-O-N-CO composition matching that of the interstellar medium rather
than winds from the  hydrogen-depleted atmosphere of a carbon
star.  The cometary globules are too massive and too dense to
match the Rayleigh-Taylor instability model.  The maximum density increase
due to a shock wave is a factor of six: less from
Rayleigh-Taylor instabilities.  Turbulence dispersion of nonlinear thin
shell instabilities
\citep{vis94} should decrease or prevent shock induced or
gravitational increases in density.  The measured density of the Helix
cometary globule atmospheres is
$\approx 10^{-14}$ kg m$^{-3}$ and their masses are a few times $10^{25}$
kg.  No mechanism is known by which such massive dense objects can form
or exist near the central star, or be ejected without disruption to the
distances where they are observed.

\subsubsection{The HGT model}

According to HGT, stars are formed by accretion of PFP planets and the
larger Jovian PFP planets (JPPs) and red dwarf stars accreted from PFPs
within the PGC interstellar medium.  The accretion mechanism is likely to be
binary, where two PFPs or  Jovians experience a near collision so that
frictional heating of their atmospheres produces evaporation of the solid
planets and an increase in the amount of gas in their atmospheres.  This
results in a non-linear ``frictional hardening'' of the binary planets
until the two objects merge, and explains why ``3 out of every 2 stars is
a binary'' (Cecilia Payne-Gaposchkin).  Heating from the merger will result
in a large atmosphere for the double-mass PFP that will increase its cross
section for capture of more PFPs.  The large atmosphere also increases the
friction with randomly encountered ambient PFP atmospheres that will slow
the relative motion of the objects and increase the time between their
collisions and mergers.  Radiation to outer space will cause the PFP
atmospheres to cool and eventually rain out and freeze if no further
captures occur, leading to a new state of metastable equilibrium with the
ambient gas.  To reach Jupiter mass,
$10^{-6} M_{\sun}$ mass PFPs and their growing sons and daughters must pair
10 times ($2^{10} \approx 10^{3}$). To reach stellar mass, 20 PFP binary
pairings are required ($2^{20} \approx 10^{6}$).  Because of the binary
nature of PFP structure formation, it seems likely that double stars will
result, as observed, and that the stars will have large numbers of large
gassy planets, as observed. Rocky planets like the Earth in this scenario
are simply the rocky cores of Jupiters that have processed the dust
accumulated gravitationally from supernova remnants in their cores and in
the cores of the thousands of PFPs that they have accreted to achieve their
masses.  Without PFPs, the existence of rocks is a mystery.

Large gas planets from PFP accretion cascades may form gently
over long periods, with ample time at every stage for
their atmospheres to readjust with ambient conditions and return to
metastable states of random motion.  These are probably the conditions under
which the old globular star clusters (OGCs) in the halo of the Milky Way
Galaxy formed their small long-lived stars.  However, if the PFP accretional
cascade is forced by radiation or tidal forces from passing stars or ambient
supernovae, a more rapid cascade will occur where the PFP atmospheres become
large and the relative motions become highly turbulent.  The turbulence
will mix the PFPs and their large planet descendants and inhibit large
average density increases or decreases.  In this case another instability
becomes possible; that is, if the turbulence weakens the creation of large
central density structures, but enhances accretion to form large planets and
brown dwarfs, the increased densities can become so rapid that buoyancy
forces may develop from the density gradients.  This will suddenly damp the
turbulence at the Schwarz turbulence scale $L_{ST}$ (see Table 1) to produce
fossil turbulence
\citep{gib99} in a volume containing many solar masses of gas, PFPs and
JPPs.  

Turbulence fossilization due to buoyancy creates a
gravitational collapse to the center of the resulting non-turbulent gas and
PFPs from the sudden lack of turbulence resistance.  A fossil turbulence
hole in the ISM will be left with size determined by the turbulence levels
existing at the beginning of fossilization.  The accretion of the planets
and gas within the hole will be accelerated by the rapidly increasing
density.  The total mass of the stars produced will be the size of the hole
times the ISM density.  If the mass is many solar masses then the
superstars formed will soon explode as supernovae, triggering a sequence of
ambient PFP evaporations, accretional cascades, and a  starburst that may
consume the entire dark PGC and PFPs to produce a million stars and a young
globular cluster (YGC) or a super-star cluster \citep{tra03}.  Numerous YCCs
are triggered into formation by galaxy mergers, such as the merging of two
galaxies and some fragments revealing a 130 kpc ($4
\times 10^{21}$ m) radius baryonic dark matter halo in the  VV29abcdef
Tadpole complex imaged by HST/ACS immediately after installation of the ACS
camera.  Figure 1 shows a dark dwarf galaxy, revealed in the baryonic dark
matter halo of the central Tadpole galaxy VV29a by a dense trail of YGCs
pointing precisely to the spiral star wake produced as the dwarf blue
galaxy VV29c and companions VV29def merged with VV29a \citep{gs03a}.  

Planetary nebula form when one of the smaller stars formed by PFP accretion
uses up all its H-He fuel to form a dense carbon core with temperature less
than $8 \times 10^{8}$ K, the star forms a white dwarf and the high
temperatures and radiation pressure of the white dwarf expel the atmosphere
of the final asymptotic giant branch red giant star (eg: Betelgeuse).  Red
giant stars have envelope diameters
$\approx 10^{12}$ m and atmospheric densities $\approx 10^{-17}$ kg m$^{-3}$
with a
$6 \times 10^6$ m diameter
$\rho \approx 10^{10}$ kg m$^{-3}$ carbon star core
\citep{cha01}, so the total mass expelled is only $\approx 10^{16}$ kg, or
$\approx 10^{-8} M_{\sun}$, much less than the gas mass values $(3-1)
\times  M_{\sun}$ observed in planetary nebulae.  Without HGT, this much
gas is mysterious.

Because the white dwarf is likely to have a companion star \citep{gue01}, it
is likely that the companion will donate mass to the white dwarf.  The white
dwarf has small size and high density, and is likely to spin rapidly with a
strong magnetic field near its spin axis.  The spinning magnetic field
lines at the white dwarf poles capture the incoming plasma to produce
powerful plasma jets perpendicular to the plane of rotation of the two
stars. The accretion disk of the white dwarf may shield some of its
radiation and broadly beam its radiation.  When the plasma jet encounters a
frozen PFP or Jupiter at the edge of the star accretion hole it  increases
the evaporation rate of the frozen gas planet and forms a cometary globule
with an outward radial wake and inward ionization front, as observed.

\section{Observations}

Figure 2 shows a mosaic of nine HST/ACS images from the F658N filter
(H$_{\alpha}$ and N II) that enhances the  ionized cometary
globules and their hydrogen tails.  See
http: //archive.stsci.edu/hst/helix/images.html.    A sphere with radius
$3
\times 10^{15}$ m is shown corresponding to the volume of
primordial-fog-particles (PFPs) with mass density $2 \times 10^{-17}$ kg
m$^{-3}$ required to form two central solar mass stars by accretion of
PFPs.  The large comets closest to the central stars must be evaporating
massive planets (Jupiters) to survive measured evaporation rates of
$2 \times 10^{-8} M_{\sun}$  year$^{-1}$
\citep{mea98} for the 20,000 year age of Helix.  Massive planets are
formed in the accretional cascade of PFPs to form stars according to HGT. 
The younger (2,000 year old) planetary nebula Spirograph (IC 418) shown
below shows
shock wave patterns from the supersonic stellar winds but no cometary PFP
candidates within its fossil turbulence accretion sphere.  The extended
spherical nebular shell for Helix contains 
$\approx 20 M_{\sun}$ of dark PFPs, from which $ 1.5
\times M_{\sun}$ has been evaporated as gas and dust
\citep{spe02}.
Evidence for bipolar beamed radiation is shown by the brighter regions of
the nebula in Fig. 2 at angles 10 and 4 o'clock, and by the light to dark
transition after 11:30 suggesting the bipolar beam is rotating slowly
clockwise.  Note that the tails of the comets are long ($\approx 10^{15}$
m) before 11:30 and short or nonexistent afterward.  Rayleigh-Taylor
instability as a mechanism to produce the globules 
\citep{cap73} gives densities much too
low.  The beam appears to have started rotation at about 1 o'clock
with deep penetration of the radiation on both sides, revolved
once, and is observed with its bright edge at 11:30 having completed less
than two revolutions to form the Helix spiral.

Figure 3 shows a Hubble Space Telescope  Helix WFPC2 1996 image to the
northeast in Helix where the closest comets to the center are found
\citep{ode96}.  The large cometary globules shown have size about
$10^{13}$ m and measured atmospheric mass $3 \times 10^{25}$ kg, with
spacing $\approx 10^{14}$ m, as expected for gas planets with some multiple
of this mass in a relic concentration corresponding to the primordial plasma
density  
$(3-1)
\times 10^{-17}$ kg m$^{-3}$ at the time of first
structure 30,000 years after the big bang.  Most of these largest cometary
globules probably have larger mass planets than PFPs at their cores to have
survived the 20,000 year lifetime of the Helix planetary nebula with
measured mass loss rates of order
$10^{-8} M_{\sun}$ year$^{-1}$ \citep{mea98}.  These are termed Jovian PFP
planets (JPPs).  The spacing of the cometary knots becomes closer for
distances farther from the central stars, consistent with these objects
having PFPs or small JPPs at their cores.

Figure 4 shows an example from the new ACS/WFPC composite images from the
northern region of Helix confirming the uniform density of the cometary
globules throughout the nebula, and presumably reflecting the ambient dark
frozen PFPs and JPPs in the surrounding ISM.  Larger JPPs
are more likely to be found near the edge of the star accretional
cavity, although the intervening PFPs at the edge would have evaporated.

Figure 5 shows the Eskimo planetary nebula, which is about twice the
distance as Helix but is still close enough for the numerous cometary
globules to be resolved.  The PNe is smaller than Helix and has a central
shocked region with no comets like the young Spirograph nebula shown at the
bottom of Fig. 2. Gas planets with primordial plasma
densities appear to be a generic feature of the ISM of this region of the
Galaxy disk within a kpc of earth where PNe are resolved clearly by HST
cameras.   

Figure 6 shows details of the central region of the Dumbbell planetary
nebula featuring numerous cometary globules and knots.  The spacing of the
objects is consistent with PFP and JPP planets with average mass density
thousands of times the average for the Galaxy dominating the mass and
species content of the ISM, and close to the primordial baryonic density of
the time of first structure in the plasma epoch 30,000 years after the big
bang as predicted by HGT.

\section{Conclusions}

High resolution wide angle HST/ACS images confirm and extend the previous
WFPC2 HST picture of the Helix planetary nebula \citep{ode96} showing
thousands of mysterious, closely spaced cometary globules with atmospheric
masses ($\approx 2 \times 10^{25}$ kg) larger than Earth-mass ($6 \times
10^{24}$ kg), comets larger than the solar system out to Pluto
($10^{13}$ m), and globule atmospheric densities ($\approx 10^{-14}$
kg m$^{-3}$) a thousand times greater than the atmosphere of the red giant
star before its expulsion to form the
planetary nebula.  The standard model of planetary nebula formation
\citep{cap73}, where 
Rayleigh-Taylor instabilities of a denser outer shell are triggered by
collision with a high speed inner shell to form the globules, cannot
account for the globules and cannot account for the large observed density
and  mass of the nebular gas and globules.   The idea that accretional
shocks can trigger gravitational instabilities
\citep{vis94} to achieve such densities neglects turbulence dispersion, and
no mechanism exists to expel such objects from the central stars.  Shocks
are seen in younger PNe than Helix (Figures 2 and 5) but are not
accompanied by any cometary globules.

We conclude that a better model for interpreting the observations is that
provided by hydro-gravitational theory (HGT) where the brightest cometary
globules are Jovian accretions of primordial-fog-particle (PFP)  Mars-mass
frozen H-He  planets formed at the plasma to gas transition 300,000 years
after the big bang \citep{gib96}, consistent with quasar microlensing
observations showing the lens galaxy mass is dominated by rogue planets
``likely to be the missing mass'' \citep{sch96}.  From HGT and all
observations, these planetary PFPs and JPPs dominate the mass and gasses of
the Galaxy disk interstellar medium (ISM) and the inner halo mass of
galaxies within a radius from the core of about 100 kpc. 
Proto-galaxies formed during the plasma epoch fragmented after transition
to gas at primordial Jeans and Schwarz scales (Table 1) to form
proto-globular-star-cluster (PGC) clouds of PFPs that comprise the baryonic
dark matter of the ISM and the inner galaxy halo.  From HST Helix images,
the density of the Galaxy disk ISM is that of proto-superclusters formed
30,000 years after the big bang; that is, 
$\rho
\approx 2
\times  10^{-17}$ kg/$\rm m^3$, preserved as a hydrodynamic fossil and
revealed by the
$(10-4) \times 10^{13}$ m separations of the PFP candidates (cometary
globules) observed in Helix that imply this density.

HST images of other nearby planetary nebula support our interpretation. 
Cometary globules brought out of the dark by beamed
radiation from a central white dwarf and companion star appears to be a
generic feature of planetary nebulae, showing that the disk ISM is dominated
by small frozen accreting planets with such  small separations that the mass
density is that of a PGC, which is $\approx 10^{4}$ larger than that of the
Galaxy.  

\acknowledgments

Most of the information in this paper would not be available without the
heroic work and dedication of astronaut John Mace Grunsfeld whose
amazing preparation and skills exhibited during the fourth space telescope
repair mission made HST/ACS images possible.

\clearpage

\begin{figure}
        \epsscale{1}
        \plotone{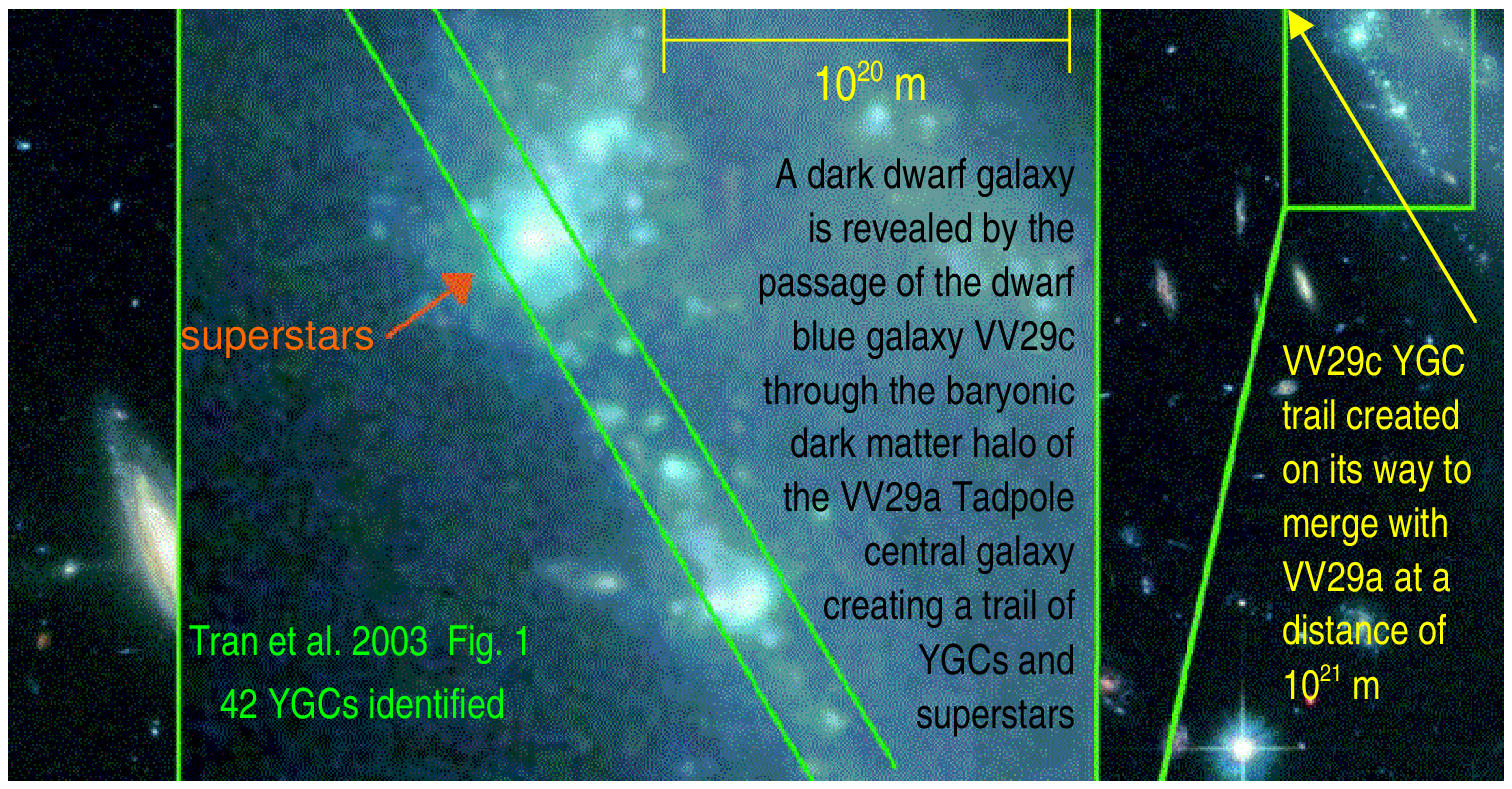}
        %%\plottwo{epsfile}{epsfile}
        \caption{Trail of 42 YGCs in a dark dwarf galaxy examined
spectroscopically by Tran et al. 2003 using the Keck telescope.  The 1 ''
Echellette slit and a loose super-star-cluster
(arrow) are shown at the left.  Ages of the YGCs range
from 3-10 Myr.  The aligned YGC trail is extended by several more YGGs
(arrow on right) and points precisely to the beginning, at $10^{21}$ m
distance, of the spiral star wake of VV29c in its capture by VV29a.  The
baryonic dark matter halo of Tadpole is revealed by a looser trail of YGCs
extending to a radius of $4
\times 10^{21}$ m from VV29a, or 130 kpc \citep{gs03a}.}
       \end{figure}

\begin{figure}
        \epsscale{0.8}
        \plotone{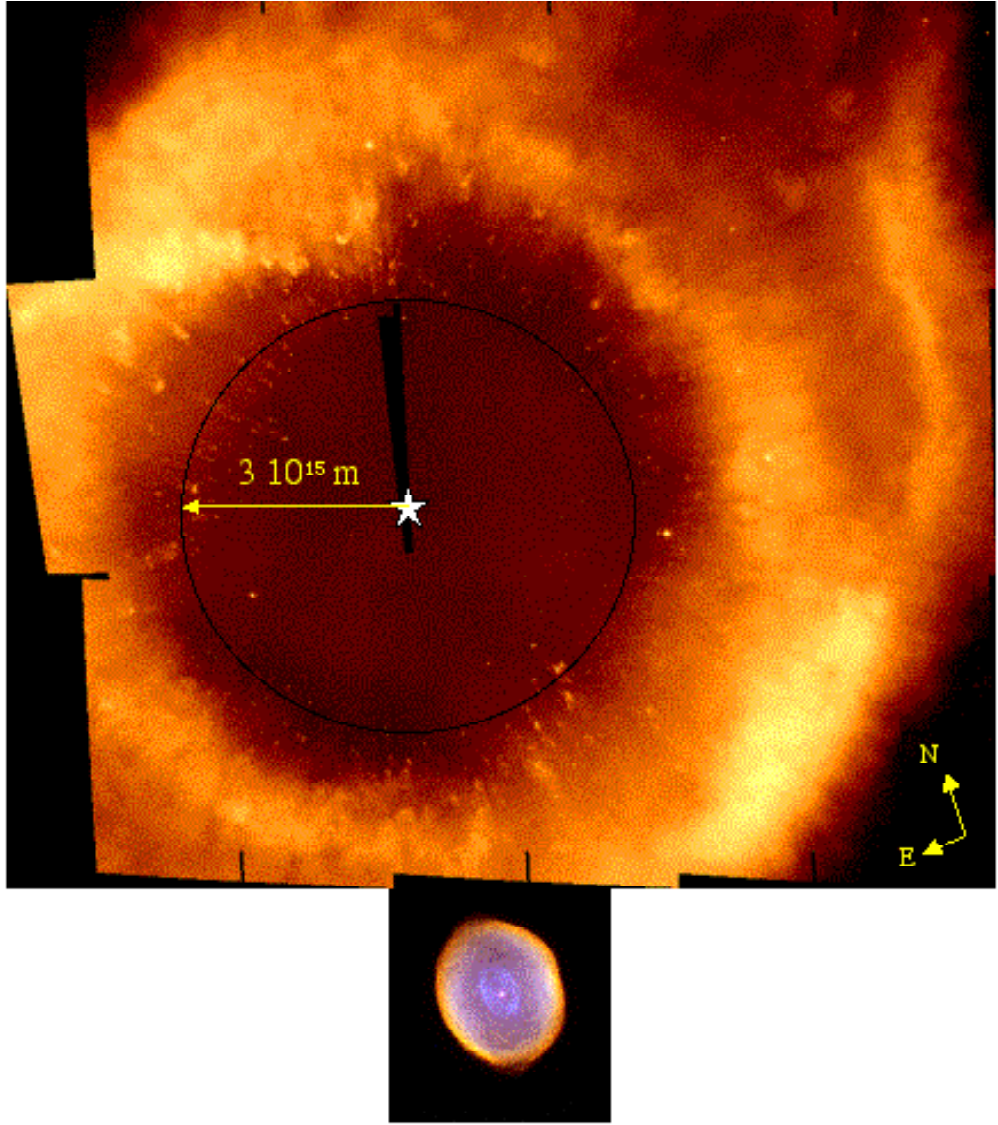}
        %%\plottwo{epsfile}{epsfile}
        \caption{Helix Planetary Nebula HST/ACS/WFC F658N image mosaic. 
The sphere has radius $3 \times 10^{15}$ m corresponding to the volume of
primordial-fog-particles (PFPs) with mass density $3 \times 10^{-17}$ kg
m$^{-3}$ required to form two central stars by accretion.  The comets
within the sphere are from large gas planets (Jupiters, JPPs) that have
survived evaporation rates of $2 \times 10^{-8} M_{\sun}$/year
\citep{mea98} for the 20,000 year age of Helix.  The younger planetary
nebula Spirograph (IC 418) shown below with no PFPs is within its accretion
sphere.  The nebular sphere for Helix contains 
$\approx 20 M_{\sun}$ of dark PFP and JPP planets, from which $ 1.5
\times M_{\sun}$ has been evaporated as detectable  gas and dust
\citep{spe02}.}
       \end{figure}

\begin{figure}
        \epsscale{.9}
        \plotone{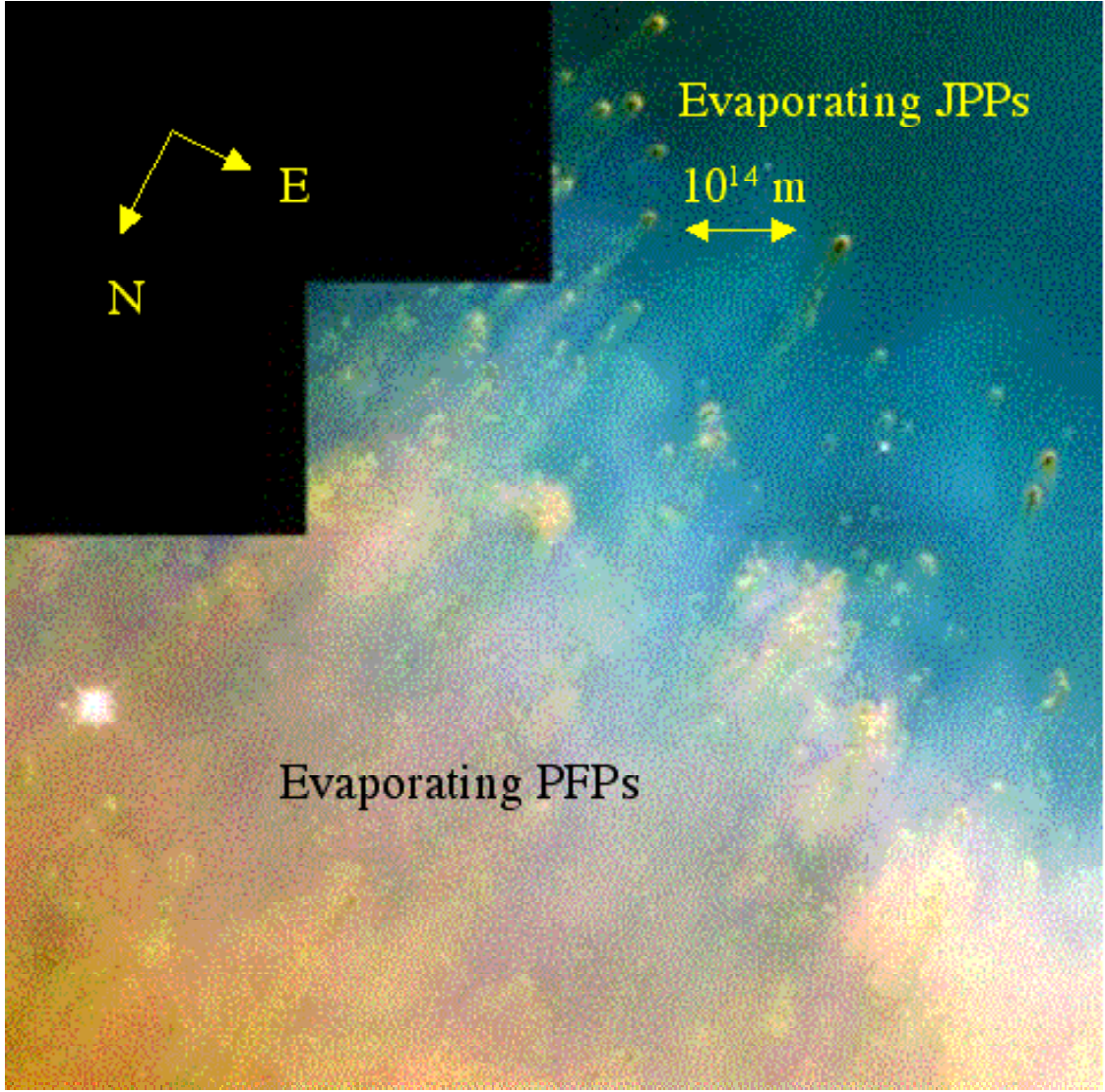}
        %%\plottwo{epsfile}{epsfile}
        \caption{Helix Planetary Nebula HST/WFPC2 1996 image \citep{ode96}
from the strongly illuminated northeast region of Helix containing the
comets closest to the central stars ($\approx 2 \times 10^{15}$ m) with
embedded $\ge 0.4$ Jupiter mass planets.  Cometary globules shown have size
about $10^{13}$ m and mass $3 \times 10^{25}$ kg just in their gas
atmospheres, with spacing
$\approx 10^{14}$ m, as expected for Jupiter mass JPPs with the
primordial density 
$3
\times 10^{-17}$ kg m$^{-3}$.  Closer spaced smaller objects without
detectable wakes are the PFPs from which the JPPs formed by accretion.}
       \end{figure}

\begin{figure}
        \epsscale{.8}
        \plotone{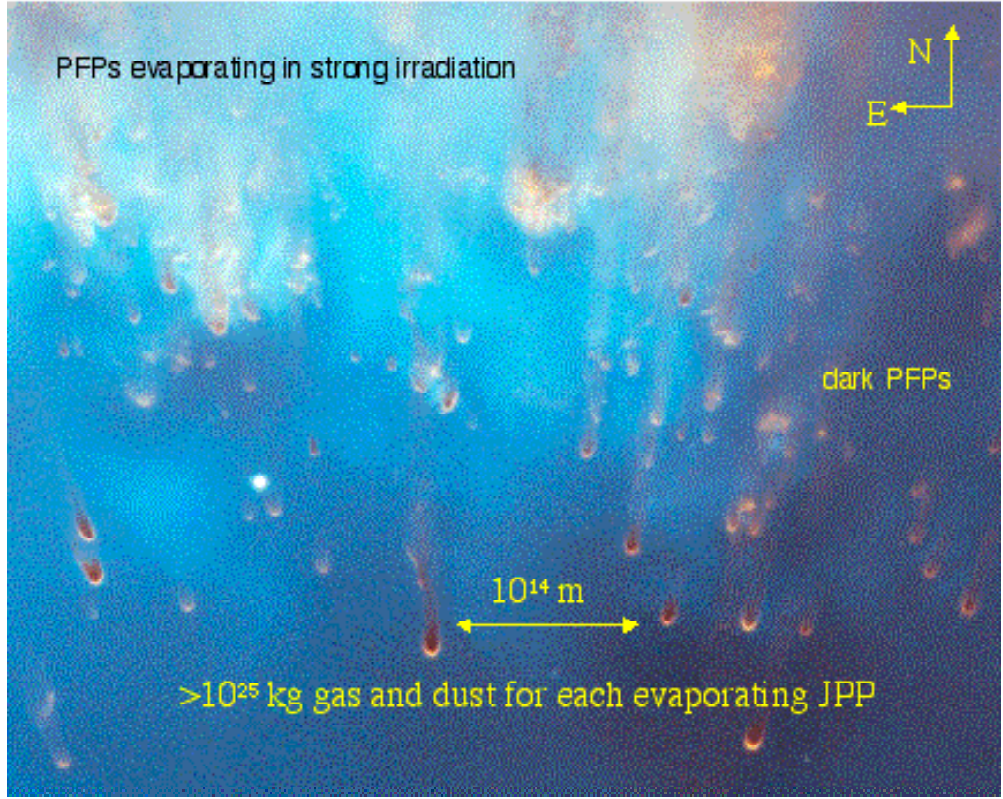}
        %%\plottwo{epsfile}{epsfile}
        \caption{Detail of more closely spaced cometary globules to the
north in Helix from the 2002 HST/ACS images at the dark to light transition
marking the clockwise rotation of the beamed radiation from the central
stars.  The comets in the dark region to the right have shorter tails and
appear smaller in diameter since they have recently had less intense
radiation than the comets on the left.  Two puffs of gas illuminated deep in
the dark region show the gas has been reabsorbed on the PFPs by gravity
during the several thousand years since their last time of strong
irradiation.  Small comets appear in the more strongly illuminated region
to the left at the larger distances from the central stars even though
these PFPs have been shielded by evaporated gas and dust.  Such PFPs are
invisible on the right in the dark region, showing they have reverted to
their original dark matter state.}
       \end{figure}

\begin{figure}
        \epsscale{.9}
        \plotone{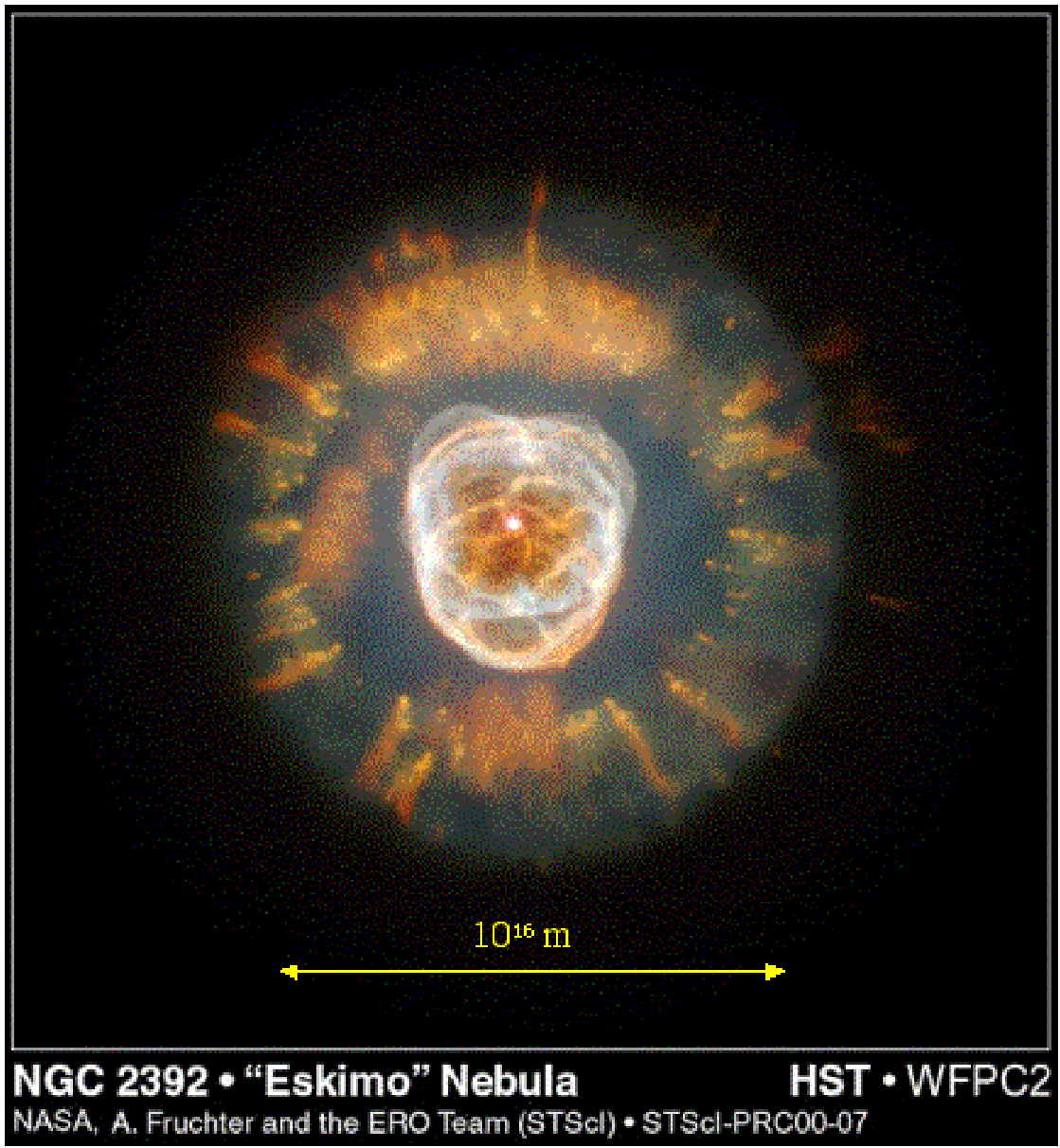}
        %%\plottwo{epsfile}{epsfile}
        \caption{The Eskimo planetary nebula (NGC 2392) is about twice the
distance from earth as Helix, but still shows numerous evaporating PFP and
JPP candidates in its surrounding interstellar medium in the HST/WFPC
images, with spacing consistent with the primordial plasma density.  The
nebula is smaller and younger than Helix, with a central shocked region
like that of Spirograph in Fig. 2.  Note the homogeneous distribution of
comets in the gas nebula and the occasional comets appearing to the right
outside the gas nebula, contrary to their formation by Rayleigh-Taylor
fragmentation \citep{cap73}. }
       \end{figure}

\begin{figure}
        \epsscale{.8}
        \plotone{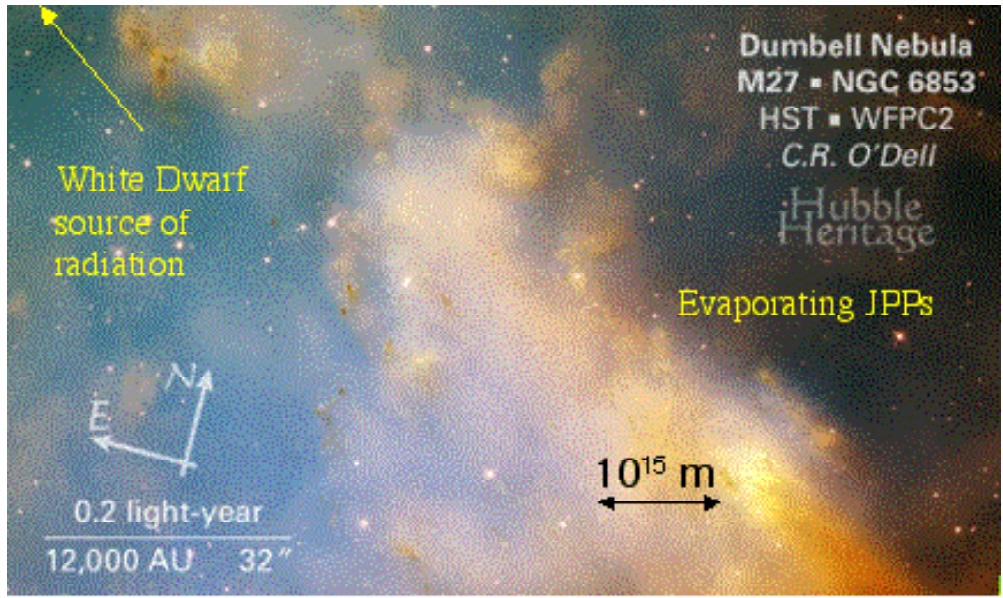}
        %%\plottwo{epsfile}{epsfile}
        \caption{Closup image of the Dumbbell planetary nebula (M27, NGC
6853) shows numerous closely spaced, evaporating, irradiated PFP and JPP
candidates in its central region.  The PNe is at a distance
$\approx 10^{19}$ m, with diameter $\approx 2 \times 10^{16}$ m.  The white
dwarf central star 
appears to have a companion from the double beamed radiation emitted.
The lack an accretional hole may be the result of
a different viewing angle (edgewise to the binary star plane) than the
face-on views of Helix (Fig. 2) and Eskimo (Fig. 5).}
       \end{figure}

\clearpage

\begin{deluxetable}{lrrrrcrrrrr}
\tablewidth{0pt}
\tablecaption{Length scales of self-gravitational structure formation}
\tablehead{
\colhead{Length scale name}& \colhead{Symbol}           &
\colhead{Definition$^a$}      &
\colhead{Physical significance$^b$}           }
\startdata Jeans Acoustic & $L_J$ &$V_S /[\rho G]^{1/2}$& ideal gas pressure
equilibration

\\Jeans Hydrostatic & $L_{JHS}$ &$[p/\rho^2 G]^{1/2}$& hydrostatic pressure
equilibration

\\ Schwarz Diffusive & $L_{SD}$&$[D^2 /\rho G]^{1/4}$& $V_D$ balances
$V_{G}$
\\  Schwarz Viscous & $L_{SV}$&$[\gamma \nu /\rho G]^{1/2}$& viscous force
balances gravitational force
  \\ Schwarz Turbulent & $L_{ST}$&$\varepsilon ^{1/2}/ [\rho G]^{3/4}$&
turbulence force  balances gravitational force
\\

Kolmogorov Viscous & $L_{K}$&$ [\nu ^3/ \varepsilon]^{1/4}$& turbulence
force  balances viscous force
\\

Ozmidov Buoyancy & $L_{R}$&$[\varepsilon/N^3]^{1/2}$& buoyancy force
balances turbulence force
\\

Particle Collision & $L_{C}$&$ m \sigma ^{-1} \rho ^{-1}$& distance between
particle collisions
\\

Hubble Horizon & $L_{H}$&$ ct$& maximum scale of causal connection
\\

%\cutinhead{This is a cut-in head}
%\sidehead{I am a side head:}

\enddata
\tablenotetext{a}{$V_S$ is sound speed, $\rho$ is density, $G$ is Newton's
constant, $D$ is the diffusivity, $V_D \equiv D/L$ is the diffusive velocity
at scale $L$, $V_G \equiv L[\rho G]^{1/2}$ is the gravitational velocity,
$\gamma$ is the strain rate,
$\nu$ is the kinematic viscosity,
$\varepsilon$ is the viscous dissipation rate, $N \equiv
[g\rho^{-1}\partial\rho/\partial z]^{1/2}$ is the stratification frequency,
$g$ is self-gravitational acceleration, $z$ is in the opposite direction
(up),
$m$ is the particle mass,
$\sigma$ is the collision cross section,  $c$ is light speed, $t$ is the
age of universe.}

\tablenotetext{b}{Magnetic and other forces (besides viscous and turbulence)
are negligible for the epoch of primordial self-gravitational structure
formation
\citep{gib96}.}

%% You can append references to a table using the \tablerefs command.

%\tablerefs{}
\end{deluxetable}

\clearpage

\begin{deluxetable}{lrrrrcrrrrr}
\tablewidth{0pt}
\tablecaption{Acronyms}
\tablehead{
\colhead{Acronym}& \colhead{Meaning}           &

\colhead{Physical significance}           }
\startdata

BDM & Baryonic Dark Matter&PGC clumps of PFPs from HGT
\\

CDM & Cold Dark Matter& questioned concept
\\

CDMHCC & CDM HCC&questioned concepts
\\

HCC & Hierarchical Clustering Cosmology& questioned concept
\\

HCG & Hickson Compact Galaxy Cluster& Stephan's Quintet (SQ=HGC 92)
\\

HGT & Hydro-Gravitational Theory& corrects Jeans 1902
\\

ISM &Inter-Stellar Medium& mostly PFPs and gas from PFPs
\\

JPP&Jovian PFP Planet&planet formed by PFP accretion
\\

NBDM & Non-Baryonic Dark Matter&includes and may be neutrinos
  \\

OGC & Old Globular star Cluster& PGC forms stars at $t
\approx 10^{6}$ yr
\\

PFP&Primordial Fog Particle&planet-mass protogalaxy fragment
\\

PGC & Proto-Globular star Cluster& Jeans-mass protogalaxy fragment
\\

SSC & Super-Star Cluster&  a cluster of YGCs
\\

YGC & Young Globular star Cluster& PGC forms stars at $t
\approx$ now
\\

%\cutinhead{This is a cut-in head}
%\sidehead{I am a side head:}

\enddata

%% You can append references to a table using the \tablerefs command.

%\tablerefs{}
\end{deluxetable}

\end{document}